\documentclass{article}

\usepackage{arxiv}
\usepackage{mathtools}
\usepackage[utf8]{inputenc} 
\usepackage[T1]{fontenc}    
\usepackage{hyperref}       
\usepackage{url}            
\usepackage{booktabs}       
\usepackage{amsfonts}       
\usepackage{nicefrac}       
\usepackage{microtype}      
\usepackage{lipsum}		
\usepackage{graphicx}
\usepackage{natbib}
\usepackage{doi}
\usepackage{amsmath}
\usepackage{amssymb}
\usepackage{multicol}
\usepackage{multirow}
\usepackage{subcaption}
\usepackage{float}
\usepackage[linesnumbered,ruled,vlined]{algorithm2e}
\usepackage{algorithmic}
\usepackage{caption}
\title{Malware Triage Approach using a Task Memory based on Meta-Transfer Learning
Framework}

\author{
  Jinting Zhu, Julian Jang-Jaccard, \\
  Cybersecurity Lab \\
  Massey University  \\
  Auckland, New Zealand\\
  \texttt{\{Jinting Zhu\}jzhu3@massey.ac.nz} \\
   \And
 Ian Welch, Harith Al-Sahaf \\
 School of Engineering and Computer Science \\
  Victoria University of Wellington, \\
 Wellington, New Zealand\\
 \\
     \And
 Seyit Camtepe \\
  ‡CSIRO Data61, \\
 Australia\\
 \\
}

\renewcommand{\shorttitle}{\textit{arXiv} Template}

\hypersetup{
pdftitle={A template for the arxiv style},
pdfsubject={q-bio.NC, q-bio.QM},
pdfauthor={David S.~Hippocampus, Elias D.~Striatum},
pdfkeywords={First keyword, Second keyword, More},
}

\begin{document}
\maketitle

\begin{abstract}
	To enhance the efficiency of incident response triage operations, it is not cost-effective to defend all systems equally in a complex cyber environment. Instead, prioritizing the defense of critical functionality and the most vulnerable systems is desirable. Threat intelligence is crucial for guiding Security Operations Center (SOC) analysts' focus toward specific system activity and provides the primary contextual foundation for interpreting security alerts. This paper explores novel approaches for improving incident response triage operations, including dealing with attacks and zero-day malware. This solution for rapid prioritization of different malware have been raised to formulate fast response plans to minimize socioeconomic damage from the massive growth of malware attacks in recent years, it can also be extended to other incident response. We propose a malware triage approach that can rapidly classify and prioritize different malware classes to address this concern. We  utilize a pre-trained ResNet18 network based on Siamese Neural Network (SNN) to reduce the biases in weights and parameters. Furthermore, our approach incorporates external task memory to retain the task information of previously encountered examples. This helps to transfer experience to new samples and reduces computational costs, without requiring backpropagation on external memory. Evaluation results indicate that the classification aspect of our proposed method surpasses other similar classification techniques in terms of performance. This new triage strategy based on task memory with meta-learning evaluates the level of similarity matching across malware classes to identify any risky and unknown malware (e.g., zero-day attacks) so that a defense of those that support critical functionality can be conducted.
\end{abstract}


\section{Introduction}
In security operations, threat intelligence is crucial in directing SOC analysts toward specific system activity and provides the background context for interpreting security alerts. Lower-threat incidents should be assigned to lower-level analysts to optimize human resources, while senior analysts should focus on more serious threats. Additionally, automated resolution should be the preferred approach for the most  dangerous malicious attacks, with manual intervention reserved only for rare cases such as unknown malware. According to the report  \footnote{https://www.enisa.europa.eu/publications/enisa-threat-landscape-2022} by the European Union Agency for Cybersecurity in 2022, which mentioned that malware and attacks against availability ranked the highest during the reporting period, malware accounted for the rank most elevated threat. Moreover, malware refers to a malicious program designed to block access to a computer system until money is paid, hence the name malware. Compared to other types of malware, the emergence and rapid increase of malware has been due to bitcoin and encryption technology. The invention of bitcoin has provided an anonymous payment channel for criminals demanding malware. The wide use of solid encryption techniques in many applications also allowed the creation of malware where decryption is impossible without knowing the cryptographic key. The rapid increase in malware has been reported causing the loss of millions of dollars for businesses and individuals. Primarily, resourceful threat actors have utilized 0-day exploits to achieve their operational and strategic  goals. The more organizations increase the maturity of their defenses and cybersecurity programs, the more they increase the cost for adversaries, driving them to develop and/or buy  0-day exploits since in-depth defense strategies reduce the availability of exploitable  vulnerabilities. However, statistic-based methods such as clustering \cite{zhu2021joint}, entropy analysis \cite{mcintosh2019inadequacy, mcintosh2018large}, similarity analysis \cite{Jang11}, information flow analysis \cite{Mirzaei17}, and examining manifest file \cite{rasthofer15} have been proposed to recognize malware quickly. However, the heavy reliance on manual analysis and tools support became either too expensive or impossible due to the growing volume of malware attacks to millions within a brief period.

The proliferation of machine learning techniques has allowed the reduction of manual intervention and has offered more automation-based machine analysis to rapidly recognize different types of malware (and other malware) to reduce a significantly increasing loss of money and productivity \cite{zhu2020multi, wei2021ae, xu2021improving, liu2021artificial}. Semi-automated approaches using random forest \cite{Laurenza20, Laurenza17} were proposed to detect malware rapidly. For example, a KNN-based \cite{Kirat13} and decision tree-based \cite{amira21} classifiers to triage a large number of malware samples rapidly have been proposed in recent years. 

However, many hurdles remain in providing effective machine learning-based triage solutions. We use “classification” to classify malware samples into specific known profiles of existing malware classes. Extending from the classification, we use the word “triage” to refer to the overall assessment process, classifying malware samples into known malware classes and identifying new types of malware samples, for example, risky and unknown malware.

One big obstacle to developing machine learning-based triage solutions is the availability of malware samples to train machine learning models to learn about the features involved in malware codes. Unfortunately, current machine learning techniques demand an extensive dataset (e.g., hundreds and thousands) to train a machine learning technique to recognize important/relevant features, detect different malware signatures, or find correlations across various elements. The other obstacle is feature representation, when the quality of parts feeding to machine learning models heavily decides the quality of outcomes (e.g., detection accuracy). In many cases, malware authors can still easily avoid detection by applying obfuscation techniques to change feature representations easily even after a machine learning model is trained with a set feature representation.

To address these issues, we propose a novel malware triage approach, and the main contributions of our proposed approach follow.

\begin{itemize}


\item We have created a novel meta-transfer learning framework for the incident response that leverages external information to optimize the learning process, enhancing the overall efficiency of the triage system. Our work marks the first attempt to utilize task-driven meta-transfer learning for malware triage, demonstrating the potential of this approach to strengthen cybersecurity operations.


\item We employ a first-order approach in Model-Agnostic Meta-Learning (MAML) \cite{finn2017model} along with fine-tuning ResNet18 to write the support items to the memory module. This simplifies and accelerates episode training without the need for backpropagation.


\item Our method employs a weighted ratio that relies on similarity scores to triage malware classes that are considered risky or unknown, such as those that exhibit features typical of malware but do not match known malware profiles exactly. This prioritizes potentially zero-day attacks for further analysis, enabling the development of appropriate response strategies.

\item We conduct extensive experiments on two public datasets for few-shot learning to demonstrate that our method can effectively leverage limited data and achieve competitive results compared with the benchmark performance.

\end{itemize}

The rest of the paper is organized as follows. Section 2 presents related works. Section 3 provides the details of the proposed approach. Section 4 describes the experiments' details and discusses the proposed approach's results, discussion, and limitations. Finally, Section 5 draws a conclusion and describes the planned future work.

\section{Related work}\label{sec:rw}

\subsection{Triage Approach}
	
Triage aims to assess the severity of the malware based on its impact on the victim. The literature offers a variety of methods to speed up the severity of classification. The aim is to automatically investigate malware samples and pass them to the proper channels (e.g., cybersecurity professionals) for further analysis. A semi-automatic malware analysis architecture proposed in \cite{Laurenza20} replaces multi-class classification with a group of one-class classifiers to decrease the runtime. A random forest classifier is used in \cite{Laurenza17} with static malware features. Static features can be extracted fast without running the malware code and help perform quick triage. However, a weakness random forest algorithm as a classifier, this approach tends to overfit when there is malware noise (e.g., obfuscated code) intentionally introduced by malware authors \cite{alsouda2019iot}.

BitShred \cite{Jang11} is a system developed for large-scale malware clustering and similarity analysis. BitShred uses feature hashing to reduce feature space dimensionality and uncovers family relations by investigating correlated features with Jaccard similarity.  SigMal \cite{Kirat13} extracts features from the malware executable headers. SigMal transforms these features into a  digital image. Signatures are extracted from images and investigated with similarity measures using the KNN algorithm. However, the KNN algorithm is susceptible to class imbalance, as pointed out by \cite{kurniawati2018adaptive}. TriDroid \cite{amira21} extracts coarse-grained features (e.g., permissions) to allocate an investigated android app into a three-class queue and uses fine-grained static features with three-class classifiers to confirm the queue assignments with high accuracy.

TRIFLOW \cite{Mirzaei17} uses information flow in Android apps to characterize behaviors of the risky apps for triage purposes. DROIDSEARCH \cite{rasthofer15} is a search engine with triage capability that triages based upon information such as ratings, downloads, or information from the manifest file. Mobile Application Security Triage (MAST) \cite{chakradeo2013mast} uses statistical methods called Multiple Correspondence Analysis (MCA), to find correlations in qualitative data obtained from the market. Rather than inspecting malware itself, tweet messages are investigated in \cite{Vinayakumar19} using Deep Neural Networks (DNN) for malware severity classification.

In the previous malware triage work, the proposed models have not achieved either satisfactory accuracy or were extensively tested on different attack classes. This leaves the triage system vulnerable only to classifying known malware samples with a high false-positive rate but also unable to identify unknown samples. The significance of the triage system is to support the automated operation of the system with less human intervention or cyber security analysts to recognize the urgency of the malicious attack to formulate more informed response strategies. Therefore, it is essential for a triage system to be equipped not only accurately classify them, but in some special cases, it identifies risky unknown malware for further analysis.

\subsection{Transfer-Learning based meta Learning}

The training strategy based on  Transfer-learning differs from the episodic training strategy in meta-learning. Instead, the conventional techniques are able to be applied in a pre-train model with a large amount of data from the base classes. And then, the pre-trained model is adapted to recognize the base classes and novel classes. \cite{yuan2020transfer,sun2019meta,bronskill2021memory,sun2020meta,soh2020meta}. Yuan et al.\cite{yuan2020transfer} proposed an offline adaptive learning algorithm that is able to be learned through the meta-training part and the fine-tuning part. The meta-training part aims to optimize the task procedure, and fine-tuning part adjusts the pre-trained parameters with the limited data in the new application task. Sun et al. \cite{sun2019meta} learned a base learner that could be adjusted to the new task with a few labeled samples.
In order to improve the efficacy of performance, they further introduced the hard task meta-batch scheme as a learning curriculum. Soh et al. \cite{soh2020meta} exploit both external and internal instances to present a Meta-Transfer Learning for the zero-shot task.


\begin{figure*}[h]
		\centering
		\includegraphics[width=1.0\textwidth]{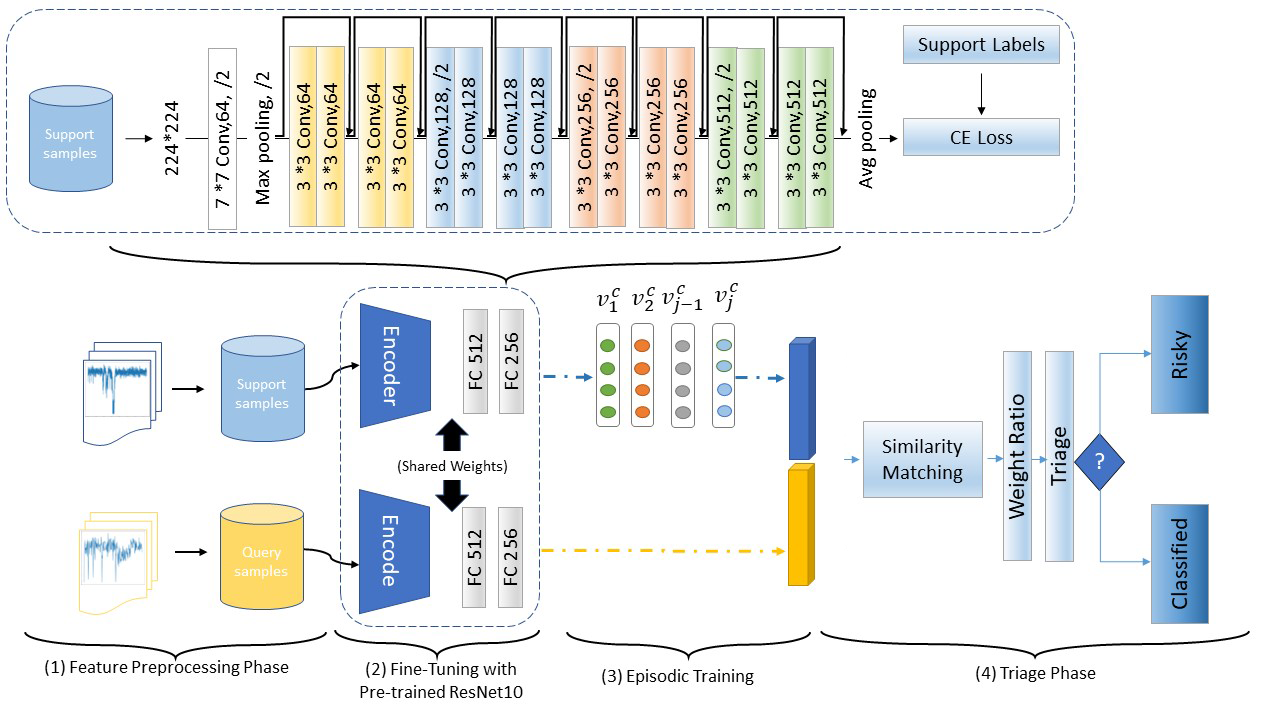}
		\caption{Episodic Training for our Triage System}
		\label{fig:overview}
\end{figure*}

\section{Methodology}\label{sec:mp}

The overview of our proposed model is shown in Fig. \ref{fig:overview}. Our model consists of three phases, (1) feature preprocessing phase, (2) meta learning phase, and (3) triage phase, respectively. The main goal of the feature preprocessing phase is to obtain entropy features from the bytecode of malware and preprocess them to feed to the model of Resnet 18 for fine-tuning. In the meta-learning phase, our model constructs the feature embedding space of the input by utilizing two identical ResNet18 networks. We load a pre-trained version of the ResNet18 network trained on more than a million images from the ImageNet database \footnote{ImageNet. http://www.image-net.org}. This pre-trained ResNet18 is used in a meta-learning fashion, where the memory features are used for the feature generalization ability.  The triage phase provides an operation service that classifies different malware classes and identifies risky and unknown malware based on the weight ratio.

\subsection{Problem Formulation}
We proceed to introduce the problem setup and notations of meta-transfer learning, which is a novel learning method that helps deep neural networks converge faster while reducing the probability to overfit when using a few labeled training data only. To train a  meta-learning model, which can be divided into few-shot learning or zero-shot learning. In this paper, we construct the few-shot tasks based on the episode training, which is defined as  an $N$-way $K$-shot classification problem. In few-shot classification, a small support set $S$ and query set $Q$ are associated with the Meta-training and Meta-testing phase. During the optimizing stage with the loss function, a parameter $\theta$ is learned from episode training data $ $. The number of episodic optimization equals the number of updating times. During the testing stage, an optimized meta-learning space measures the correlation between unseen samples in the support $S$ and query set $Q$.

\begin{algorithm} [th!]
	\SetKwInOut{Input}{Input}
	\SetKwInOut{Output}{Output}
	
	\Input{$f$: malware binary file; $l$: segment length; $n$: the number of files }
	\Output{entropy pattern image $m$}
	
	\While{not reach to n}{
		1. read $l$ bytes from $f$, and define as a segment $s_i$;
		
		2. \textbf{for} $j = 0$ to $255$ \textbf{do}
		
		\ \ \   \ \ \ \ \ \ 2.1 compute the probability $p_j$ of $j$ appearing in $s$ using the Shannon entropy function
		
		3. an entropy graph image  $m$ generated from whole segments in a binary file\
	}
	
	
	\caption{Generating Entropy Features}
	\label{alg:entropy_graph}
\end{algorithm}

\subsection{Feature Preprocessing Phase}
Entropy is often used as the measure of changes in information content. More changes to the original information content produce higher entropy values, while fewer changes to the original information are associated with lower entropy values. As discussed in \cite{zhu2022few, zhu2021task}, using entropy values as feature representation has several advantages compared to using grayscale image features. The feature values associated with grayscale features can be easily fooled if malware authors apply certain obfuscation techniques. For instance, control flow obfuscation techniques \cite{wang2021efficient,azar2021data,chow2001approach} can easily alter the control flow path of a malware program. For example, it inserts a junk code or shuffles the order of function calls which does not affect the semantics of the original malware program. This simple technique can effectively cause the appearance of grayscale images to be different from each other. This can easily lead to misclassification results where a classification algorithm puts them into different malware families when in fact, they belong to the same malware family. In addition, the computation cost associated with grayscale image features is usually much higher due to the higher cost involved in processing texture features of grayscale images \cite{zhu2022few, zhu2021task}. To avoid these weaknesses associated with grayscale features, we instead use entropy features that are more resilient to changes in the malware binary code and reduce computation costs.

We first read a malware binary file as a stream of bytes to construct entropy features. The bytes are made into multiple segments, each comprising 200 bytes. We further count the frequency of unique byte values in the range of pixel values between 0 and 255, followed by computing the entropy using Shannon’s formula as seen in the following Eq. (\ref{eq:entropy}).
\begin{equation}
    H(p)=-\sum_{i} p_ilogp_i
    \label{eq:entropy}
\end{equation}
where $p_i$ is the probability of an occurrence of a byte value. The entropy obtains the minimum value of 0 when all the byte values in a binary file are the same, while the maximum entropy value of 8 is obtained when all the byte values are different. The entropy values are then concatenated as a stream of values and are used to generate an image-based entropy graph. The entropy graph, an image of size 224*224, is fed as entropy features as input to our model. Generating the entropy features from a malware binary file is depicted in Algorithm \ref{alg:entropy_graph}.

\subsection{Resnet 18 with Fine Tuning for Feature Extraction }
Figure. \ref{fig:overview} shows a typical parameter-level Fine-Tuning (FT) the operation, which is in the meta-optimization phase. At the first stage, we fine-tune Resnet 18 with the support set to adjust the feature embedding, and then it is prepared to be applied in the backbone network to generate feature embedding for the meta-learning phase (i.e., the entropy feature representing a malware byte code from the same family). The structure of the ResNet18 network follows the paper except for the fully connection layers (FC),  in which the first fully connected layer has 512 neurons and the last fully connected layer serves as the output layer and has 256 neurons. The last fully connected layer from each Resnet18 is combined together to create a single fully connected layer with 512 neurons to compute a loss function across the features processed by two ResNet18 networks. The inputs are fed by the entropy graph that is represented as an image form of  2-dimensional vectors of a fixed size 224 $\times$ 224,  the fine-tuned ResNet18 is further optimized to obtain updated weights and parameters. The four blocks of each ResNet18 are followed by two fully connected layers with a fattening layer in between.




.

\subsection{Meta-Transfer learning}

The first-order for optimization  follows the MAML algorithm, which aims  \cite{finn2017model} to shallow the gap from the distribution of unseen samples. Specifically, the weights of feature extractor $\Theta$ is fine-tuned with other datasets, and the original classifier $\theta$ is replaced with an FC layer we designed, as the data set for fine-tuning  and then optimize them by Adam optimizer as follows:
\begin{equation}
    [\Theta;\theta] = : [\Theta;\theta] -a\nabla\mathcal{L}_{\mathcal{D}}([\Theta;\theta])
\end{equation}
where $\mathcal{L}$ denotes the following empirical loss,

When the fine-tuned model is obtained, we begin the training process for meta transfer-learning outlined with a set of weights optimized for the last Residual Network block.  This makes the last residual network block that can easily adopt weights to new examples, thus ensuring effective handling of novel inputs. Despite being trained solely on the malware source domain, the malimage dataset has been observed to facilitate the acquisition of transferable features during training, as noted in \cite{sun2019meta}.  Moreover, transferring statistics is used in this paper to calibrate the distribution of a limited number of training samples, which can be challenging as the learned model is prone to overfitting due to the biased and limited number of samples. The specific training process is outlined in Algorithm \ref{alg:training}.

During the meta-learning phase, the classifier $\theta$ will be discarded, because subsequent few-shot tasks contain different classification objectives, such as the N-way classification.  To further optimize  the meta operations scaling and shifting through meta-batching training. Given a task $\mathcal{T}$,  the loss of $\mathcal{L}^{tr}$ is used to optimize the current base-learner (classifier) $\theta$ by gradient descent.

\begin{equation}
    \theta_i = \theta - \alpha \nabla_{\theta}\mathcal{L}^{tr}_{\mathcal{T}_i}(\theta)
\end{equation}
where $\alpha$ is the task-level learning rate,  the loss of $\mathcal{T}$ is used to optimize the current base-learning (classifier) $\theta^{'}$ by gradient descent:

\begin{algorithm} [th!]
	\SetKwInOut{Input}{Input}
	\SetKwInOut{Output}{Output}
	
	\Input{ Restnet18 (Pre) ; Support Set $\mathcal{S}$ and Query Set $\mathcal{Q}$ in Base Set }
	\Output{updated $\theta_i$}
	\textbf{Require:} Model parameter $\theta_N$ trained for  base learner; \\
	Support sample in task $\mathcal{T}_i$;  \\
	Query batch in task $\mathcal{T}_i$\\
	\While{not reach in episodes of $\mathcal{T}_{train}$)}{
		Randomly sample task \( \mathcal{T}_i \) \\
		$x_s,x_q$ = Feature extractor $\Theta$ \\
		Evaluate $\mathcal{L}_D( ([\Theta; \theta^{(meta)}]))$ = $d(h_s, h_q)$  by Eq. 2;\\
		Optimize the  $\Theta$ and  $\theta^{(meta)}$ by Eq.10  and   Adam \\
	} 
	\caption{Learning in Meta-Training Phase}
	\label{alg:training}
\end{algorithm}
\begin{algorithm} [th!]
	\SetKwInOut{Input}{Input}
	\SetKwInOut{Output}{Output}
	
	\Input{Feature Extractor Learned: $\Theta$; Classifier $\theta^{meta}$; A new task Support $\mathcal{S}$ and Query set $\mathcal{Q}$ in Novel Set}
	\Output{Predicted Label of $N$-way $y_i$}
	\textbf{Require}: Task Memory: 		$v_m = \frac{1}{N_c}\sum_{i=1}^{N_c}f(x_i^j;\Theta)$ \\
	\While{not reach to episodes in $\mathcal{T}_{test}$}{
		
		accuracy = $p(h_{s,q},v_m;\theta^{(meta)})$  \\

	}
	avg\_acc = running\_acc / test\_episode \\ 
	\caption{Prediction in  Meta-Testing Phase}
	\label{alg:entropy_graph}
\end{algorithm}
\subsection{Task Memory for Adapted Semantic Feature}

We make an external memory module \cite{zhang2021learning} that aims to store the previous experience from the fine-tuning stage with true label $y_m$. Each of the columns represents as average features of one class from other support sets. Memory columns given by, which could be queried by finding the top $k$ neighbors between a feature and memory columns. Meanwhile, the feature in the support and query set could absorb the task-specific information and therefore able to adopt the current task.

Inspired by the concept we mentioned above, our aim is to create a semantic embedding invariance  refer to the prototypical network \cite{snell2017prototypical}  and in which the feature vector $\mathbf{v_i}$ can be constructed to a memory module called task-memory. We define the entries in  $\mathbf{V}$ = $\{\mathbf{v_1},\mathbf{v_2}, \mathbf{v_3},..., \mathbf{v_k}\}$, in which each entries $\mathbf{v_i}$ is an average feature performance computed via  all instances in one class of the support set, and which could be represented as:
\begin{equation}
    v_j = \frac{1}{n}\sum_{j=1}^{n_i}
    f(x_i^j;\Theta)
\end{equation}
where $f(\cdot)$ is a feature embedding constructed with $\Theta$ and input feature $\mathbf{x_i^j}$ ,  where $\mathbf{v_j}$ is the memory feature calculated from the support set in the  base set and $\mathbf{x_i}$ is an instance of total number $n_i$  in the feature embedding of base set. Gao et al. \cite{gao2019hybrid} argued that a given instance may has a higher variance from the mean feature calculated by all of rest feature or some of them have been wrongly labeled. Moreover, the small number of sample tend to result in a large variance so that a straightly average performance may not estimate a truly distribution over the all instances. To enhance the ability of feature representation, we make use of the extent memory to compute the relation between the support and query samples, which correlation score  $a_{j,i}$  could be computed based on softmax function.

\begin{equation} \label{eq:co_dis}
  Sim( f(\mathbf{x_i^j};\Theta), \mathbf{v_j})  = \frac{ f(\mathbf{x_i^j};\Theta) \cdot v_j}{|| f(\mathbf{x_i^j};\Theta) ||||  \mathbf{v_j}||}
\end{equation}
which is used to compute a weights, $a_j$, computed by a softmax function. And a memory vector could be retrieved by this weight and weighted average on the task specific information of support set and query set according to follow equation:

\begin{equation}
    a_j = \frac{exp(Sim (f(\mathbf{x_i^j};\Theta), \mathbf{v_j} )}{\sum_{k=1}^{n_i}exp(Sim (f(\mathbf{x_i^j};\Theta), \mathbf{v_k} ))}
\end{equation}
where \textit{tanh} is selected as  an activation function before the softmax function to produce weights $a_j$ for the inner product of the two feature vectors of instances. And then, we aggregate the sequentially arriving semantic feature $\mathbf{x_i^j}$ regardless of whose gradient and then average all the instances embedded in the support set for each relation. After reshaping the feature maps, we feed them into the memory module to obtain the corresponding feature vectors. To determine the memory size $m$, we set it equal to the size of the support set, and which feature vectors contained with task-information could be represented as with corresponding weights: 

\begin{equation}
   \mathbf{v_m} = \sum_{j=1}^{n_i}a_j\mathbf{v_j^i}
\end{equation}
where $\mathbf{v_{m}}$ is the adaptive prototype for a class $i$  and $a_{j}$ indicates the class weights on each memory cluster. so that a high similar score given by the relation are obtained of higher weights to the instance in the support set.

\begin{equation}
   \mathbf{h} = \tau \mathbf{v_m} + (1-\tau)f(f(\mathbf{x_i^j};\Theta) )
\end{equation}
where $\tau$ is the hyper-parameter determined by cross validation. To construct training episodes, a random selection of classes is made from the training set, followed by selecting a subset of examples within each class to form the support set.

\begin{equation}
    p_j^e= \frac{exp (-d( \mathbf{h_{s_i}}, \mathbf{h_{q_i}})) )}{exp (-d(\sum_{c^i \in V^c}exp( \mathbf{h_{s_i}},\mathbf{h_{q_i}} ))} 
\end{equation}
where $d(\cdot, \cdot)$ is the Euclidean distance function  for relation vectors according to the strategy of Snell et al. \cite{snell2017prototypical} and it also generates a distribution over the specific class with the cross entropy loss function.

 \begin{figure*}[h]
		\centering
		\includegraphics[width=0.9\textwidth]{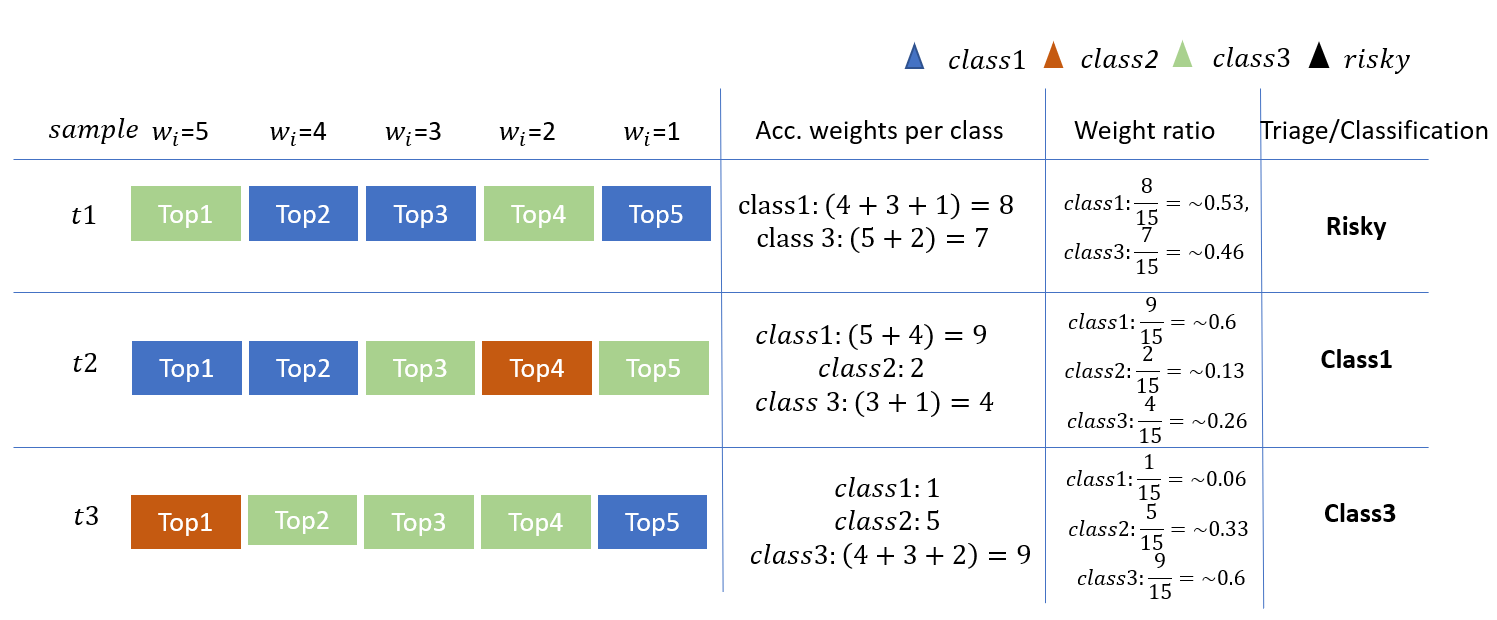}
		\caption{Weight Ratio and Triage}
		\label{fig:weight_ratio}
\end{figure*}

\subsection{Triage Phase}
Our triage approach is similar to one used in the hospital system. Like patients whose symptoms match the known profiles of illnesses, any (test) malware samples whose features match the known malware classes are classified immediately to formulate a rapid response strategy. In some patient diagnosis whose symptoms does not clearly match one known profile but several profiles of illnesses, they could be sent to a designated specialist for further examinations. Our triage also recognizes this type of malware sample whose feature clearly demonstrates that it is malware but does not match a specific malware family. These cases are classified as risky and unknown malware so that further analysis is carried out rapidly to establish if this is a new variant (e.g., zero-day attack). Towards this approach, our model first uses similarity matching to search if a malware sample under the examination (i.e., test sample) exhibits features similar to any known malware classes. The level of similarity matching to different malware classes is computed as a weight ratio. By evaluating the weight ratio, we can triage a malware sample to if this can be classified as specific malware or should be classified as risky unknown malware.


\subsubsection{Similarity Matching}
In a general similarity searching approach, a query record is compared against a stored database of records. The main goal here is to retrieve a set of database records that are similar to the query record. For example, if there is a picture of a dog as a query record, a similarity search should give a list of pictures with dogs in them.

In the context of the triage approach, the database corresponds to a collection of vectors of malware, which is a collection of feature embedding trained for a pair of training samples from a malware class.
 Similarity searching in our context refers to searching for similar feature embeddings. We employ the FAISS algorithm \footnote{https://github.com/facebookresearch/faiss/blob/main/INSTALL.md}, Facebook’s library for faster similarity searching even for very large datasets, to calculate the similarity based on cosine distance. The details of the cosine distance we use are:


In our model, we use the FAISS algorithm \cite{guo2003knn,ma2021pod} to return the top $K$ most similar feature embeddings of malware samples whose cosine distance between the feature embedding of a test sample and the feature embeddings of the training sample are minima. Here, a similarity score indicates a different relationship between two samples (i.e., a test sample and a class representative from the trained malware classes). The minimum value of 0 indicates two samples are completely dissimilar. The maximum value of 1 indicates two samples are completely similar. The value close to 0 indicates the characteristics of orthogonality or decorrelation, while in-between values indicate intermediate similarity or dissimilarity.

\subsubsection{Weight Ratio and Classification}
Based on the top $K$ search results and their weights, a weight ratio is computed and the final tirage/classification is done. Mathematically, weight $w_i$ to the $i$th nearest neighbor for $i=1,..., K $ and classifies the test sample $x$ as the class that is assigned the most weight $w_i$, as follows,
\begin{equation}
\label{eq:weights_ratio}
    W_{nor} = argmax \sum_{i=1}^{K}w_iI_\{y_i=g\}
\end{equation}
where  $w_i$ denotes that the weights and $I$ is the total weights in the same class.

We normalize and regularize the weight ratio in a style that satisfies a linear interpolation with the maximum entropy (LiME) objective by linearly combining each weight \cite{van2012kernel}. This allows the combined weights to be better balanced and the test sample, $x$,  is best approximated based on the training samples through the Equation \ref{eq:dictionary_weight}. The LiME objective is described as follows:
\begin{equation}
\begin{aligned}
 &\min_{d}=|| \sum_{i=1}^k d_ix_i-f(t)||_2^2- \lambda ||\sum_{i=1}^kd_i||_2^2 \\
 & subject \ \  to \sum_{i=1}^k d_i=1, d_i \geq 0, i = 1, ..., k
\end{aligned}
\label{eq:dictionary_weight}
\end{equation}
where $x_i \rightarrow \mathbf{R}$ is $i$th training sample, the $d_i$ is the basis atoms for the feature vector, $x_i$, and $f(t)  \rightarrow \mathbf{R}$ is the feature vector of test sample, $t$, $\lambda$ is a regularization parameter. 

 Fig. \ref{fig:weight_ratio} illustrates how the weight ratios are computed to decide a triage/classification outcome. Let’s take a test sample $t_1$ whose top 5 similar results were from the malware class 3 and malware class 1. The weights from the malware class 3 were 5 (had the highest similarity score based on FAISS similarity matching) in the top 1 position and 2 (had the 2nd lowest similarity score), respectively. This results in the total accumulated weight for the malware class 3 = 7. The weight ratio is calculated by dividing the total accumulated weights for a class by the total weights of search results which results in a weight ratio of approximately 0.53 for class 1 and 0.46 for class 3. Let’s presume that the weight ratio rate we want to decide to classify a given sample is set at the threshold = 0.6. Neither the weight ratio of class1 nor class3 satisfies this threshold. In this case, our model will put the test sample in the “risky and unknown” class and the sample does not exhibit any clear features that can classify into a known malware class despite it exhibiting some common features from known malware classes. In contrast, take a test sample  $t$  where the weight ratio for class1 is equal to (or greater than) the threshold. In this case, the $t$ is classified as belonging to the malware class1.



\section{Experimental Results}\label{sec:er}
This section describes the details of our experiments including the system environment, dataset, and performance metrics we used. We also discuss the experimental results with discussion.  
\subsection{Dataset}

\begin{table}[h]
		\caption{Details of the malware dataset}
		\label{tbl:malware_dataset}
		\centering
		{\footnotesize
			\begin{tabular*}{0.48\textwidth}{@{\extracolsep{\fill}}lcc}
				\toprule
				
			\textbf{Class Name}&	\textbf{Instances} & \textbf{Ratio} (\%) 
				\\
				\midrule
				Bitman & 99 & 9.45 
				\\
				Cerber & 91 & 8.68 
				\\
				Dalexis & 9 & 0.86 
				\\
				Gandcrab & 100 & 9.54 
				\\
				Locky & 96 & 9.16 
				\\
				Petya & 6 & 0.57 
				\\
				Teslacrypt & 91 & 8.68 
				\\
				Upatre & 18 & 1.72 
				\\
				Virlock & 162 & 15.46 
				\\
				Wannacry & 178 & 16.98 
				\\
			Zerber & 198 & 18.89 
				\\
				\bottomrule	
			\end{tabular*}
		}
\end{table}

\begin{table}[h!]

  \caption{Details of the Malimage dataset}
  \centering
		{\footnotesize
			\begin{tabular*}{0.48\textwidth}{@{\extracolsep{\fill}}lcc}
				
			\toprule
			\textbf{Class Name} & \textbf{Family} & \textbf{Instances} 
				\\ \midrule
				
				Worm & Allaple.L & 1591 
				\\
				Worm & Allaple.A & 2949 
				\\
				Worm & Yuner.A & 800 
				\\
				PWS & Lolyda.AA 1 & 213 
				\\
				PWS & Lolyda.AA 2 & 184 
				\\
				PWS &   Lolyda.AA 3 & 123 
				\\
				Trojan & C2Lop.P & 146 
				\\
				Trojan & C2Lop.gen!g  & 200 
				\\
				Dialer & Instantaccess & 431 
				\\
				TDownloader & Swizzot.gen!I & 132 
				\\
			TDownloader &  Swizzor.gen!E & 128 
				\\
				Worm & VB.AT & 408 
				\\
				Rogue & Fakerean & 381 
				\\
				Trojan & Alueron.gen!J & 198 
				\\
				Trojan & Malex.gen!J & 136 
				\\
					PWS & Lolyda.AT & 159 
				\\
				Dialer & Adialer.C & 125 
				\\
					TDownlaoder & Wintrim.BX & 97 
				\\
					Dilaer & Dialplatform.B & 177 
				\\
					TDownlaoder &  Dontovo.A & 162 
				\\	TDownlaoder &  Obfuscator.AD & 142 
				\\	Backdoor & Agent.FYI & 116 
				\\
					Worm:AutoIT & Autorun.K & 106 
				\\
					Backdoor & Rbot!gen & 158 
				\\
					Trojan & Skintrim.N & 80 
				\\
						\bottomrule
				
 \end{tabular*}
\label{table:malimage}
		}
\end{table}

We created a dataset containing malware binaries from ViruseShare \footnote{VirusShare. https://virusshare.com/}. The dataset comprises a total number of 1,048 samples from 11 families/classes of malware, each of which consists of a varying number of samples as listed in Table \ref{tbl:malware_dataset}. The distribution of data in the VUW dataset accurately reflects real-world situations, with some classes, such as Petya and Dalexis, being outnumbered by other classes like Zerber, as indicated in the third column of Table I.

In contrast, the Malimage dataset, as described in \cite{Nataraj2011} and listed in Table II, consists of 9,314 instances from 25 different malware families. The width and height of each malware image differ between families, but Nataraj et al. \cite{Nataraj2011} standardized the width based on the file size and bytecode sequence. The number of samples in the Malimage dataset exceeds that of the VUW dataset. Additionally, the Malimage dataset features distinct image textures among the various malware families.

\subsection{Experimental Environment}
\begin{table}[h]
    	\caption{System configuration}
    	\label{table:Mat}
    	\centering
    	{\footnotesize
    		\begin{tabular*}{0.45\textwidth}{@{\extracolsep{\fill}}lp{8cm}}
    			\toprule
    			\textbf{Unit}   & \textbf{Description}\\ 
    			\midrule
    			Processor   & 3.6 GHz 8-core Inter Core i7 \\ 
    			RAM  &  32 GB      \\
    			GPU  & GeForce GTX 2070      \\
    			Operating System  & Windows 10  \\	
    			Packages   &  Pytorch, Sckit-Learn, Numpy, Pandas   \\ 
    	
    			\bottomrule
    		\end{tabular*}
    	}
\end{table}
This study was carried out using a 3.6 GHz 8-core Intel Core i7 processor with 32 GB memory on Windows 10 operating system. The proposed approach is developed using Python programming language with several statistical and visualization packages such as Sckit-learn, Numpy, Pandas, and Pytorch. Table~\ref{table:Mat} summarizes the system configuration for our environment.



Table \ref{table:Training parameters} illustrates the best hyperparameters we used for our proposed model, as well as the  1-shot for 19 batch size;
5-shot for 15 batch size. The classes in Table I exhibit a significant variation, with nearly half of them containing no more than 25 malware samples. In fact, some classes had only one sample, which could be attributed to the recent detection of new malware types such as Blocal and Newbak. To address this issue, we employed a data augmentation technique that increased the image sample size for each malware class to a minimum of 30 samples. This technique involved applying random transformations to the images, such as rotations and re-scaling, with rotations set at angles of 90, 180, and 270 degrees. Additionally, we ensured that the mean value and standard deviation were set to 0.52206 and 0.08426, respectively. In order to evaluate the performance of our models, we conducted measurements of 1-shot and 5-shot accuracy across 2-way and 5-way classifications using random datasets drawn from the complete collection of training and test sets in 20,000 distinct episodes. It should be noted that the size of the malware images remained constant for each model throughout the evaluation process and half for training and the rest half for testing

\begin{table}[h!]
  
    \footnotesize\centering
    \caption{Training parameters}
    \label{table:Training parameters}
    
    \begin{tabular}{ccc}
        \midrule
       \textbf{  Hyperparameters} & \textbf{Values} &\textbf{ Descriptions} \\ \midrule
        Learning rate & 0.001 & Learning speed (within range 0.0 and 1.0) \\
        Batch size & 64 & No. of samples in one fwd/bwd pass \\
        Epoch & 50 & No. of one fwd/bwd pass of all samples \\\
        Memory size & k &  size of support set\\
        \bottomrule
    \end{tabular}
\end{table}

\begin{table*}[]
    \centering
 \caption{Accuracy scores (\%)  tested on two dataset}
 \setlength\doublerulesep{0.4pt}
 
\begin{tabular}{ p{4.5cm} p{2cm} p{2cm} p{2cm} p{2cm}}
\toprule[1pt]\midrule[0.3pt]
\cmidrule(lr){2-5}



\multirow{2}{*}{\textbf{Few-shot Learning Model}}& \textbf{2-way 1-shot }& \textbf{2-way 5-shot}& \textbf{5-way 1-shot }& \textbf{5-way 5-shot}\\\cline{2-5}

&\multicolumn{4}{c}{\textbf{VUW dataset}}\\


\hline
Siamese \cite{snell2017prototypical} & 71.6 + - 1.6\%  & 73.2 +- 2.0\% & 60.5 + - 1.4\% &  63.1 + - 1.9\% \\
Relation  \cite{sung2018learning}  & 55.8 + - 1.8\%& 55.2 + - 2.6\%   & 42.6 + - 3.1\%& 43.2 + - 2.1\%     \\
Prototypical (4CONV) \cite{snell2017prototypical}   & 79.3 + - 2.0\%  & 84.2 + - 1.8\% & 67.8 + - 2.0\% & 69.2 + - 1.9\%\\
Matching \cite{milosevic2017machine}   & 74.1 + - 2.2\%  &  62.2 + - 1.7\%  &   63.4 + - 2.1\% &  64.4 + - 2.5\%   \\

\textbf{Ours} (Resnet 18 (Pre))& 83.1 + - 1.5\% &  86.1 +- 1.7 \%& 73.5 +- 1.6\% &  75.8 +- 1.9\% \\
\textbf{Ours} & 84.8 + - 1.8\% &  87.5 +- 1.8 \%& 75.7 +- 2.1\% &  77.3 +- 2.3\%\\
\midrule[0.3pt]
 & \multicolumn{4}{c}{\textbf{Malimage dataset}} \\
\hline
Siamese \cite{snell2017prototypical} & 79.6 + - 2.0\%   &  83.0 + - 1.9\% &  71.3 + - 1.7\% &  73.2 + -2.1\% \\
Relation \cite{sung2018learning}  & 62.4 + - 2.5\% & 67.2 + - 2.7\%& 55.7 + - 3.2\% & 57.1 + - 2.7\%\\
Prototypical (4CONV) \cite{snell2017prototypical} & 94.6 + - 1.8\% &  95.5 + - 1.5\%& 91.1 + - 2.0\% & 95.9 + - 1.8\%\\

Matching \cite{milosevic2017machine} & 86.9 + - 2.1 & 88.3 + - 1.8\%  &  80.2 + - 2.4\%  & 81.6 + - 2.2\%   \\

\textbf{Ours} (Resnet 18 (Pre)) & 95.9 + - 2.3\% & 96.4 + - 1.7\% & 92.2 + - 1.9\% &  98.1 + - 1.9\%\\
\textbf{Ours}  &96.8  + - 1.6\% & 97.2 + - 1.8\% & 94.3 + - 2.2\% & 98.7 + - 1.7\% \\
\midrule[0.3pt]\bottomrule[1pt]
\label{results}
\end{tabular}

\label{table:results_init_trained}
\end{table*}

\subsection{Classification Performance}

\paragraph{\textbf{Vuw Dataset Results}} Table 5 demonstrates the superior few-shot classification performance of our proposed model with Task-memory and ResNet18(pre), achieving an accuracy of 84.8\% for (2-way, 1-shot). Additionally, our model's accuracy of 87.5\% for (2-way, 5-shot) is comparable to the benchmark results of 79.3\% and 84.2\% reported by Prototypical \cite{snell2017prototypical}. However, it is worth noting that Prototypical's use of ResNet18 without task memory resulted in a 1.7\% and 1.4\% decrease compared to our model in 2-way, respectively. Furthermore, it is important to highlight that models utilizing ResNet18(pre) outperformed those using 4CONV models by a significant margin. This is evidenced by the fact that our best result is 10.8\% higher than the best 1-shot result of 5.5\% reported in \cite{snell2017prototypical} using 4 CONV models.

\paragraph{\textbf{Malimage Dataset Results}}
The proposed method achieves the highest accuracy, with 94.3\% and 98.7\% on 5-way 1-shot and 5-way 5-shot, respectively, as shown in Table 5. This represents a significant improvement over previous methods. Our method's success is largely attributed to the novel feature extractor, which generates task-memory feature maps that adopt the  previous information expression. 

Our approach stands out compared to benchmarks such as Matching and Prototypical Nets, which typically use shallow networks to extract image features. In addition, unlike typical triage methods \cite{Laurenza20}, our approach employs meta-transfer learning, resulting in significantly higher accuracy rates for 5-way 1-shot and 5-way 5-shot classification. Our feature extraction network also outperforms others, with the best-performing result highlighted. By combining meta-transfer learning and task memory, our model can learn a more practical feature embedding and achieve greater convergence stability, leading to even higher accuracies, particularly in 2-way and 5-way classification tasks.




\subsection{Ablation Study}

\subsubsection{Discriminative Feature   from External Memory}

\begin{figure}
    \centering
    	\subfloat[Task memory and Resnet18 (pre)]
    	{\includegraphics[width=0.40\textwidth]{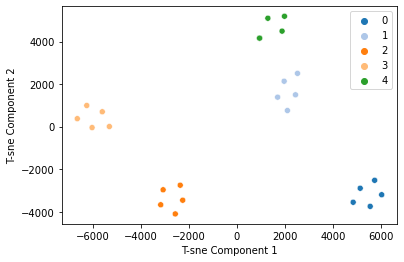}}\hfil
             \subfloat[Prototypical (4CONV)]
    	{\includegraphics[width=0.40\textwidth]{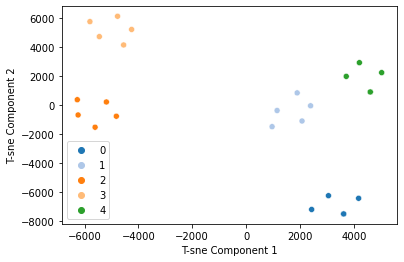}}\hfil  

        \caption{T-SNE visualization  for 5-way 5-shot testing result}
\label{t-sne-model:sum_average}
\end{figure}

In order to provide a more comprehensive understanding of the impact of discriminative learning, we employed t-SNE to create a visual representation of the feature representation extracted from our model. This is illustrated in Fig. \ref{fig:tsne}, which clearly shows that the model utilizing the task memory and Resnet18 (pre) effectively distributes the samples in the embedding space. Samples from the same classes are clustered together and those from different classes are separated, the feature extraction process is able to effectively capture highly discriminative features, which in turn leads to a high level of predictive performance.

\subsubsection{Shared Weights on Transfer learning}

By reducing the parameters of your model by half from 2$d \rightarrow d$, you have effectively decreased the number of parameters that need to be optimized. This leads to faster convergence towards a minimum during training. However, the downside of this approach is that it makes your model less flexible. It is important to note, however, that this reduction in flexibility can sometimes work as a regularizer and help to prevent overfitting. This is because the weights in the model are shared with other neurons, which can help to prevent the model from overfitting  to the training data. Observed from the Table \ref{table:results_init_trained}, the method with Resnet18 (pre) exceeds the performance of  Prototypical (4CONV) on both datasets.

This is because a common weight-sharing technique involves treating the input as a series of local regions arranged in a hierarchy. This technique is based on the prior knowledge that the input can be broken down into local regions with the same properties, and therefore each of these regions can be processed using the same set of transformations. By making this assumption, we can reduce the number of parameters required in the network as compared to a fully-connected network. This reduction in parameters can lead to an increase in the network's ability to generalize, provided that the prior assumption is correct for the problem at hand.

\section{Triage Analysis}

we could utilize the similarity compared with other features of samples from other families and put the unknown malware into the risky pool. In other words, this is able to classify the unknown and known samples with  rates into the risky pool. We further examined the sensitivity of the threshold used for classification and accuracy, which is shown in Table \ref{tab:risk_pool}. This is based on the similarity score obtained by running the KNN algorithm used in the FAISS similarity search with k size = 20. Here the threshold is the value associated with the normalized and regulized weight ratio.

\begin{table} [h!]
\centering
\begin{tabular} {c|c|c|c}
\hline
 Threshold  & Classified    &   Risk Pool  &   Accuracy    \\

\hline
$W_{nor} = 0.50 $ &  57.8\% &  42.2\%  &   93.0 \% \\

\hline
  $W_{nor}= 0.45$ & 59.2\%  & 40.8 \%  & 90.9\%   \\

\hline
  $W_{nor}= 0.40$ & 61.2\%  & 38.8 \% &  78.4\%   \\

\hline
  $W_{nor}= 0.30$ & 85.1\%  &14.9 \%  & 65.5\%  \\
\hline
\end{tabular}
\label{tab:risk_pool}
\end{table}

We take the weight ratio with higher confidence into account. Concretely speaking, when we used the highest threshold = 0.50, slightly more than half of the test samples = 57.8\%, were fully classified into known malware classes while slightly less than half were classified as risky unknown malware. With such a high threshold rate, the accuracy rate was also very high because at this stage the similar score contributing to weight ratio calculation has to be very high (i.e., two samples being compared need to have a very high correlation). As expected, as the threshold value decreased, more test samples were classified into known malware classes as it require a lower level of similarity score computed between a test sample and one classified during training. The accuracy rate dropped orthogonally compared to the decrease of the threshold value. At the lowest threshold value = 0.30, 85.1 \% of the test samples were classified into known malware classes while 14.9\% of the test samples went into the risky and unknown malware class. The lowest accuracy rate of 65.5\% was achieved at the lowest threshold rate.

\subsection{Discussion and Limitations}
In this study, we propose our model that can be used as a triage application based on the classification of malware. Instead of using image features, we utilize entropy features that are more robust against noises (e.g., changes made by obfuscation technique) and less computationally complex. We use pre-trained two ResNet18 networks in a meta-learning fashion to obtain accurate weights and other parameters when our training sample size is limited. Instead of a feature embedding created based on a single input, our model constructs a feature embedding based on two positive inputs from the twin ResNet18. This results in our feature embedding containing relevant features to detecting a certain malware class. Note that we only use positive inputs (i.e., two samples from the same malware family) to train our model to find more common features exhibited within the same malware family. This approach is different from other proposals where negative inputs (i.e., two samples from different malware families) are used \cite{chen2021exploring,wang2021fully,shao2021few}.  Our experimental results confirm that the use of positive inputs contributes to improving detection accuracy compare to the models using the mix of negative inputs. The triage part of our model utilizes weight ratio to compute a better classification result by taking into account the similarity scores produced by the feature vectors of test samples compared to the feature vectors of the whole training samples. This allows our triage system to classify known malware classes with high accuracy while being able to classify risky unknown malware. These risky unknown malware samples can be given a high priority for further analysis.

At the moment, our model simply classifies any samples that cannot be mapped into any known malware classes as the risky and unknown malware class as these samples exhibit the features of malware. However, it may require further deep analysis to investigate the true nature and severity of these samples (e.g., these may well be zero-day attack or maybe disguised through more complex obfuscation techniques based on the known malware classes). We observed that there could be a potential bias when the feature vectors from unknown samples are made into feature embeddings trained with known samples. To reduce this bias, we may require further analysis of the nature of unknown samples before feeding them to our model.

Our model classifies any unseen test samples into the risky and unknown pool. However, it is possible that some of these samples may exhibit some features that appear in malware but it can be benign. Our model can be trained with some benign samples to understand the features associated with these types of samples to more clearly classify whether it is completely benign or risky malware.  By using entropy features, our model is more resilient to producing misclassification when obfuscated malware is included in training samples. Though the resilience against the control flow obfuscation technique has been evidenced in \cite{zhu2021task}, the influence of other types of obfuscation techniques requires further investigation.

\section{Conclusion}\label{sec:conclusion}
We proposed a malware triage approach that can rapidly classify different malware classes even in the presence of unknown classes. Our Siamese Neural Network (SNN) based approach utilizes a pre-trained ResNet18 in a meta-learning fashion to generate more accurate feature embedding and overcomes the biases in weight and parameter calculation typically associated with a model trained with training samples. Instead of image features typically used as inputs to machine learning-based malware detection and triage applications, our approach use entropy features directly obtained from the malware binary files. Our evaluations confirm that the use of entropy features provide a better feature representation and contribute towards improving triage accuracy. The experimental results tested on various malware samples show a very high classification accuracy exceeding 88\%. In addition, we offer a new triage strategy that can recognize risky and unknown malware which exhibits the feature commonly seen in other known malware but exact matching profiles cannot be found. These types of malware can be easily prioritized for further analysis to formulate an appropriate response strategy faster before any significant damages emerge (e.g., the loss of ransom payments or reduced productivity).

We plan to extend our work to assign more sophisticated weights for the matching feature vectors for Top-$k$ nearest neighbors as discussed in \cite{dorfer2018end,li2018end,yates2020capreolus} and also include benign samples to compare their features with existing malware samples.

\section*{Acknowledgment}
This research is supported by the Cyber Security Research Programme—Artificial Intelligence for Automating Response to Threats from the Ministry of Business, Innovation, and Employment (MBIE) of New Zealand as a part of the Catalyst Strategy Funds under the
grant number MAUX1912.
\bibliographystyle{unsrtnat}
\bibliography{references}  






\end{document}